\documentclass[10pt]{article}

\def\CRAS{C.~R.~Acad.~Sc.~Paris}

\def \ccomma{\raise 2pt\hbox{,}} 
\def \D {\hbox{d}}
\def \Log {\mathop{\rm log}\nolimits}

\def \arg {\mathop{\rm arg}\nolimits}
\def \tr  {\mathop{\rm tr}\nolimits}
\def \det {\mathop{\rm det}\nolimits}
\def \diag{\mathop{\rm diag}\nolimits}

\def\CRAS{C.~R.~Acad.~Sc.~Paris}

\def\SAM{Stud.~Appl.~Math.~}

\def\PVI    {{\rm P6}}

\def\abcd{\alpha,\beta,\gamma,\delta}

\def\baru{\bar u}
\def\barpsi{\bar \psi}

\begin{document}

\title{
A reduction of the resonant three-wave interaction to the generic
sixth Painlev\'e equation
\footnote{Journal of Physics A, to appear, 
Special issue ``One hundred years of Painlev\'e VI''. 
nlin.SI/0604011. 
\hfill\break\noindent
Corresponding author RC.} 
}

\author{
Robert Conte\dag,
A.~Michel Grundland${}^{+}$,
and
Micheline Musette\ddag
{}\\
\\ \dag Service de physique de l'\'etat condens\'e (CNRS URA 2464),
\\ CEA--Saclay, F--91191 Gif-sur-Yvette Cedex, France
\\ E-mail:  Robert.Conte@cea.fr
{}\\
\\ + Centre de recherches math\'ematiques, Universit\'e de Montr\'eal
\\ Case postale 6128, Succursale Centre ville,
\\ Montr\'eal, Qu\'ebec H3C 3J7, Canada
\\    D\'epartement de math\'ematiques et d'informatique,
\\ Universit\'e du Qu\'ebec \`a Trois-Rivi\`eres
\\ Case postale 500, Trois-Rivi\`eres, Qu\'ebec G9A 5H7, Canada
\\ E-mail:  Grundlan@crm.umontreal.ca
{}\\
\\ \ddag Dienst Theoretische Natuurkunde, Vrije Universiteit Brussel,
\\ Pleinlaan 2, B--1050 Brussels, Belgium
\\ E-mail:  MMusette@vub.ac.be
}

\maketitle

\hfill 

{\vskip -10.0truemm}
{\vglue -10.0truemm}

\begin{abstract}
Among the reductions of the resonant three-wave interaction system
to six-dimensional differential systems,
one of them has been specifically mentioned as being linked to the generic
sixth Painlev\'e equation $\PVI$.
We derive this link explicitly,
and we establish the connection to a three-degree of freedom
Hamiltonian previously considered for $\PVI$.

\end{abstract}




\baselineskip=12truept 


\tableofcontents

\section{Introduction}

The three-wave resonant interaction system (3WRI) in 1+1 dimensions
i.e.~whose impulsions $k_j$ and pulsations $\omega_j$
have a zero sum,
$k_1+k_2+k_3=0, \omega_1+\omega_2+\omega_3=0$,
can be mathematically described
by
six coupled first order partial differential equations (PDEs)
in six dependent complex variables $u_j,\baru_j$ (the amplitudes)
and two independent variables $x,t$ \cite{ZM_Nwave1973},
\begin{eqnarray}
& & \left\lbrace
\begin{array}{ll}
\displaystyle{
    u_{j,t} + c_j     u_{j,x} - i \baru_k \baru_l=0,\
}
\\
\displaystyle{
\baru_{j,t} + c_j \baru_{j,x} + i u_k u_l=0,\
i^2=-1,
}
\end{array}
\right.
\label{eq3waves}
\end{eqnarray}
in which $(j,k,l)$ denotes any permutation of $(1,2,3)$,
$c_j$ are the constant values of the group velocities,
with $(c_2-c_3)(c_3-c_1)(c_1-c_2)\not=0$.

This system admits a third order Lax pair \cite{ZM_Nwave1973}.
In the traceless zero curvature representation,
this is given by \cite{AH} 
\begin{eqnarray}
& & \rho = -\frac{c_3-c_1}{c_2-c_3},\
\sigma = \frac{c_1-c_2}{c_3-c_1},\
\\
& & L=
 \frac{i \lambda}{c_1-c_2} \pmatrix{
-1 + 2 \rho & 0 & 0 \cr 0 & 2 - \rho & 0 \cr 0 & 0 & -1 - \rho \cr}
\nonumber \\ & & \phantom{1234}
 + \frac{i}{c_1-c_2} \pmatrix{
0 & - \sigma \rho u_3 & \sigma \rho \baru_2 \cr
 \sigma \baru_3 & 0 & - \sigma u_1 \cr
 - u_2 & - \baru_1 & 0 \cr},
\\
& & M=
   \frac{i \lambda}{c_1-c_2} \pmatrix{
c_1 -2 c_2 \rho & 0 & 0 \cr
0 & -2 c_1 + c_2 \rho & 0 \cr
0 & 0 & c_1 + c_2 \rho \cr}
\nonumber \\ & & \phantom{1234}
+ \frac{i}{c_1-c_2} \pmatrix{
0 & c_3 \sigma \rho u_3 & -c_2 \sigma \rho \baru_2 \cr
- c_3 \sigma \baru_3 & 0 & c_1 \sigma u_1 \cr
  c_2 u_2 & c_1 \baru_1 & 0 \cr},
\\
& &
\lbrack \partial_x - L,\partial_t-M\rbrack=0,
\label{eqLaxPDE}
\end{eqnarray}
in which $\lambda$, the spectral parameter, is an arbitrary complex constant.

The purpose of this paper is to show the existence of at least one
noncharacteristic one-dimensional reduction
to a system of ordinary differential equations (ODEs)
integrable with the generic sixth Painlev\'e function,
and to integrate it explicitly.
Indeed, at present time,
various reductions of this system
have been integrated with
most of the six Painlev\'e functions \cite{FLMS,Kitaev,MW},
but no explicit link with the generic sixth Painlev\'e equation
has been found up to now.

Since a noncharacteristic reduction preserves the order,
it is necessary,
in order to integrate with $\PVI$ which depends on four parameters $\abcd$,
that the reduced system of ordinary differential equations (ODEs)
depends on two arbitrary parameters.
The determination of all subgroups of the invariance group of the
3WRI system,
which allows one to generate all the classical reductions,
has been performed in Ref.~\cite{MW}.

The paper is organized as follows.
After recalling in section \ref{section_Local_singularity_analysis}
the singularity structure of the system,
we define the reduction in section \ref{sectionReduction},
then generate its first integrals from the Lax pair in section
\ref{sectionTwoFirstIntegrals}.
The explicit integration with the generic $\PVI$ equation
is performed in sections
\ref{sectionLink_to_SDIa}
and
\ref{sectionSinglevaluedness}.
In section \ref{sectionDual_Lax_pairs},
we discuss the link with two previous works on the same kind of
third order matrix Lax pair,
and the possible implications on a second order matrix Lax pair
for $\PVI$.

\section{Local singularity analysis}
\label{section_Local_singularity_analysis}

The singularity structure analysis of the 3WRI system (\ref{eq3waves})
has been performed in the more general setting
of three space variables \cite{GR1998}.
The result is a unique family of movable singularities,
in which the six components $u_j,\baru_j$ all behave like a simple pole
\begin{eqnarray}
& &
    u_j \sim a_j X^{-1},\
\baru_j \sim b_j X^{-1},\
\label{equjlocal}
\end{eqnarray}
in the neighborhood of a singular manifold \cite{WTC},
\begin{eqnarray}
& &
\varphi(x,t) - \varphi_0=0,
\label{eqPDEManifold}
\end{eqnarray}
in which $\varphi$ is an arbitrary function of the independent variables,
$\varphi_0$ an arbitrary movable constant,
and
the expansion variable $X(x,t)$ \cite{Cargese1996Conte}
vanishes along with $\varphi-\varphi_0$ and satisfies
\begin{eqnarray}
& &
X_x=1 + O(X),\
X_t=-C + O(X),\
C=-\frac{\varphi_t}{\varphi_x}.
\end{eqnarray}
The linearized system of (\ref{eq3waves}) in the neighborhood of
the expansion (\ref{equjlocal})
is of Fuchsian type near the singular manifold,
and its six Fuchs indices $r$ are $r=-1,0,0,2,2,3$.

The existence of the isospectral Lax pair (\ref{eqLaxPDE})
implies that no movable logarithms can enter
the expansion (\ref{equjlocal}),
which is indeed the case \cite{GR1998}.

For any noncharacteristic reduction,
the resulting system of ODEs also admits the above family of movable
simple poles,
and the first integrals can only have the singularity degrees $2,2,3$.

\section{A noncharacteristic reduction and its Lax pair}
\label{sectionReduction}

The problem of finding a noncharacteristic redution of the 3WRI system
and its Lax pair has been tackled by several authors
using different theoretical frameworks.
Among them,
Kitaev \cite{Kitaev} gave the one-dimensional reduction
(already given in \cite{FLMS} in the restricted case $\beta_j=0$)
\begin{eqnarray}
& &
\left\lbrace
\begin{array}{ll}
\displaystyle{
\zeta=\frac{x}{t},\ 
\beta_1+\beta_2+\beta_3=0,
}
\\
\displaystyle{
    u_j(x,t)= \left(t (c_j - \zeta)\right)^{-1 + i \beta_j} \psi_j,\
}
\\
\displaystyle{
\baru_j(x,t)= \left(t (c_j - \zeta)\right)^{-1 - i \beta_j} \barpsi_j,\
}
\end{array}
\right.
\label{eqReductionNonCharacteristic}
\end{eqnarray}
in which $\beta_j$ are constants,
to the six first order ODEs
\begin{eqnarray}
& &
\left\lbrace
\begin{array}{ll}
\displaystyle{
\frac{\D}{\D \zeta}
\psi_j= \phantom{-} i
 (c_j-\zeta)^{- i \beta_j}(c_k-\zeta)^{-1-i \beta_k}(c_l-\zeta)^{-1-i \beta_l}
\barpsi_k \barpsi_l,
}
\\
\displaystyle{
\frac{\D}{\D \zeta}
\barpsi_j= - i
 (c_j-\zeta)^{  i \beta_j}(c_k-\zeta)^{-1+i \beta_k}(c_l-\zeta)^{-1+i \beta_l}
\psi_k \psi_l.
}.
\end{array}
\right.
\label{eq3waves6dimsystem}
\end{eqnarray}
in which $(j,k,l)$ denotes any permutation of $(1,2,3)$.
However,
he performed the integration only in a particular case.

As noticed by Kitaev,
$\zeta,c_1,c_2,c_3$ only contribute by their crossratio,
so this system depends only on two parameters $\beta_j$.
It is nevertheless advisable to keep the $c_j$'s to display the ternary
symmetry.

To compute the reduced Lax pair,
let us represent the PDE Lax pair as the $1$-form
\begin{eqnarray}
& &
\omega = L \varphi \D x + M \varphi \D t,
\label{eqPDEOneform}
\end{eqnarray}
in which $(L,M)$ depends on $(x,t,\lambda)$.
One wants to find two operators
${\mathcal L}$, ${\mathcal M}$,
and one scalar variable $\mu$,
so as to represent the reduced Lax pair as
\begin{eqnarray}
& &
\Omega = {\mathcal L} \Phi \D \zeta + {\mathcal M} \Phi \D \mu,
\label{eqODEOneform}
\end{eqnarray}
in which $({\mathcal L},{\mathcal M})$ depends on $(\zeta,\mu)$.

One first eliminates $u_j,\baru_j,x, \D x$
from the reduction (\ref{eqReductionNonCharacteristic})
to obtain
\begin{eqnarray}
& &
\omega = L_1(\zeta,\lambda,t) \varphi \D \zeta
           + M_1(\zeta,\lambda,t) \varphi \D t,
\end{eqnarray}
then one applies a change of basis
\begin{eqnarray}
& &
\varphi = P \Phi,
\end{eqnarray}
in which the transition matrix $P$ is chosen to depend only on $t$,
so as to gather the dependence on $(t,\lambda)$
into a single variable
$\mu$.
This matrix takes the form
\begin{eqnarray}
& &
P= \diag(t^{i(\beta_3-\beta_2)/3},
         t^{i(\beta_1-\beta_3)/3},
         t^{i(\beta_2-\beta_1)/3}),
\\
& &
\Omega = P^{-1} \omega - P^{-1} (\D P) P^{-1} \varphi,
\end{eqnarray}
and the result is (\ref{eqODEOneform}), with $\mu = \lambda t$.

The reduced traceless Lax pair $({\mathcal L},{\mathcal M})$
in zero curvature representation
\begin{eqnarray}
& &
\lbrack \partial_\zeta - {\mathcal L},\partial_\mu-{\mathcal M}\rbrack=0,
\end{eqnarray}
depends on the constant spectral parameter $\mu$,
\begin{eqnarray}
{\mathcal L}
& = &
 \frac{i}{c_1-c_2} \mu \pmatrix{
-1 + 2 \rho & 0 & 0 \cr 0 & 2 - \rho & 0 \cr 0 & 0 & -1 - \rho \cr}
\nonumber
\\
& &
 + \frac{i}{c_1-c_2} \pmatrix{
0 &
- \sigma \rho    \psi_3 (c_3-\zeta)^{-1 + \beta_3 i} &
  \sigma \rho \barpsi_2 (c_2-\zeta)^{-1 - \beta_2 i} \cr
  \sigma      \barpsi_3 (c_3-\zeta)^{-1 - \beta_3 i} &
 0 &
- \sigma         \psi_1 (c_1-\zeta)^{-1 + \beta_1 i} \cr
-                \psi_2 (c_2-\zeta)^{-1 + \beta_2 i} &
-             \barpsi_1 (c_1-\zeta)^{-1 - \beta_1 i} &
 0 \cr},
\\
{\mathcal M}
& = &
 \frac{i}{c_1-c_2} \pmatrix{
        c_1-\zeta  - 2 \rho (c_2-\zeta) & 0 & 0 \cr
0 & -2 (c_1-\zeta) +   \rho (c_2-\zeta) & 0 \cr
0 & 0 & c_1-\zeta  +   \rho (c_2-\zeta) \cr}
\nonumber
\\
& &
+ \frac{i}{3} \mu^{-1} \pmatrix{
\beta_2-\beta_3 & 0 & 0 \cr
0 & \beta_3-\beta_1 & 0 \cr
0 & 0 & \beta_1-\beta_2 \cr
}
\nonumber
\\
& &
 - \frac{i}{c_1-c_2} \mu^{-1} \pmatrix{
0 &
- \sigma \rho    \psi_3 (c_3-\zeta)^{  \beta_3 i} &
  \sigma \rho \barpsi_2 (c_2-\zeta)^{- \beta_2 i} \cr
  \sigma      \barpsi_3 (c_3-\zeta)^{- \beta_3 i} &
0 &
- \sigma         \psi_1 (c_1-\zeta)^{  \beta_1 i} \cr
-                \psi_2 (c_2-\zeta)^{  \beta_2 i} &
-             \barpsi_1 (c_1-\zeta)^{- \beta_1 i} &
0 \cr}.
\label{eqLax3ODE}
\end{eqnarray}
The singularities of the matrix ${\mathcal M}$ in the complex
spectral parameter are $\mu=0$ (of the Fuchsian type) and $\mu=\infty$
(of the nonFuchsian type).

\section{The two first integrals, and the reduced fourth order system}
\label{sectionTwoFirstIntegrals}

The presence of one Fuchsian singularity in the monodromy matrix
${\mathcal M}$ allows one to generate easily the first integrals.
Indeed,
denoting ${\mathcal M}_{-1}$ the residue of the
matrix ${\mathcal M}$ at the Fuchsian singularity $\mu=0$,
\begin{eqnarray}
& &
{\mathcal M}={\mathcal M}_{-1} \mu^{-1} + {\mathcal M}_{0},
\label{eqMLaurent}
\end{eqnarray}
the invariants of the residue ${\mathcal M}_{-1}$ are constants of the
motion.
These are generated by the characteristic polynomial
\begin{eqnarray}
& &
\det ({\mathcal M}_{-1} - z)=
-z^3
- \left(K_1 + \frac{\beta_1^2+\beta_2^2+\beta_3^2}{6}\right) z
\nonumber
\\
& &
\phantom{123456}
+ 2 i \left(K_2
 - \frac{(\beta_2-\beta_3)(\beta_3-\beta_1)(\beta_1-\beta_2)}{54}
\right),
\label{eqcharpoly}
\end{eqnarray}
in which
$K_1,K_2$ denote the only two first integrals
\begin{eqnarray}
K_1
& = &
\left\lbrack
(c_2-c_3)\psi_1\barpsi_1+(c_3-c_1)\psi_2\barpsi_2+(c_1-c_2)\psi_3\barpsi_3
\right\rbrack
\left((c_2-c_3)(c_3-c_1)(c_1-c_2)\right)^{-1},
\\
K_2
& = &
\left\lbrack
\phantom{i}
 \frac{1}{2} \psi_1\psi_2\psi_3
    (c_1-\zeta)^{ \beta_1 i}(c_2-\zeta)^{ \beta_2 i}(c_3-\zeta)^{ \beta_3 i}
\right.
\nonumber
\\
& &
   + \frac{1}{2} \barpsi_1\barpsi_2\barpsi_3
    (c_1-\zeta)^{-\beta_1 i}(c_2-\zeta)^{-\beta_2 i}(c_3-\zeta)^{-\beta_3 i}
\nonumber
\\
& &
\left.
   + \frac{1}{6} \left((\beta_2-\beta_3)(c_2-c_3)\psi_1\barpsi_1
              +(\beta_3-\beta_1)(c_3-c_1)\psi_2\barpsi_2
              +(\beta_1-\beta_2)(c_1-c_2)\psi_3\barpsi_3
         \right)
\right\rbrack
\nonumber
\\
& &
\times \left((c_2-c_3)(c_3-c_1)(c_1-c_2)\right)^{-1}.
\label{eqK1K2}
\end{eqnarray}

These two first integrals have the singularity degrees $2$ and $3$,
in agreement with the results of section
\ref{section_Local_singularity_analysis}.
The two first integrals allow us to reduce the order from six to four.
Introducing the six variables $\rho_j,\varphi_j$,
\begin{eqnarray}
& &
   \psi_j=\rho_j e^{ i \varphi_j},\
\barpsi_j=\rho_j e^{-i \varphi_j},\
\label{eqpsij_barpsi_j}
\end{eqnarray}
the two invariants only depend on the four variables $\rho_j,\chi$,
\begin{eqnarray}
& &
K_1=
\left\lbrack
(c_2-c_3) \rho_1^2 + (c_3-c_1) \rho_2^2 + (c_1-c_2) \rho_3^2
\right\rbrack
\left((c_1-c_2)(c_2-c_3)(c_3-c_1)\right)^{-1},
\label{eqK1chirho}
\\
& &
K_2=
\left\lbrack
\rho_1 \rho_2 \rho_3 \cos \chi
   + \frac{1}{6} \sum_j (\beta_k-\beta_l)(c_k-c_l)\rho_j^2
\right\rbrack
\left((c_1-c_2)(c_2-c_3)(c_3-c_1)\right)^{-1},
\label{eqK2chirho}
\\
& &
\chi=\sum_j (\varphi_j + \beta_j \Log (c_j-\zeta)).
\end{eqnarray}
Therefore the differential system for $\rho_j,\chi$ is closed.
This allows one to discard the three variables $\varphi_j$,
remembering only their first derivatives
\begin{eqnarray}
& &
\varphi_j'=\frac{\rho_k \rho_l}{(c_k-\zeta)(c_l-\zeta)\rho_j} \cos \chi,
\label{eqSystemphi}
\end{eqnarray}
and to focus on the closed fourth order system
\begin{eqnarray}
& &
\rho_j'=\frac{\rho_k \rho_l}   {(c_k-\zeta)(c_l-\zeta)      } \sin \chi,
\label{eqrhoSystemchirho}
\\
& &
\chi'=\sum_j \left(
    \frac{\rho_k \rho_l}{(c_k-\zeta)(c_l-\zeta)\rho_j} \cos \chi
    - \frac{\beta_j}{c_j-\zeta}\right),
\label{eqchiSystemchirho}
\end{eqnarray}
which admits the two first integrals
(\ref{eqK1chirho})--(\ref{eqK2chirho}).

In order to integrate the system
(\ref{eqrhoSystemchirho})--(\ref{eqchiSystemchirho}),
it is advisable to lower the order from four to two,
for instance by building a single second order ODE depending on the four
parameters $K_1,K_2,\beta_j$.

\section{Link to a classified second order second degree ODE}
\label{sectionLink_to_SDIa}

Following the procedure of Ref.~\cite{MW},
we derive the change of variables which allows
the fourth order system
(\ref{eqrhoSystemchirho})--(\ref{eqchiSystemchirho})
to be explicitly integrated in terms of the generic $\PVI$ equation.

Given any two components $\rho_j^2$,
they admit a unique (up to a multiplicative factor) linear combination $Y$
whose first derivative has no contribution from $\sin \chi$,
e.g.~\cite[Eq.~(5.41)]{MW}
\begin{eqnarray}
& &
\left\lbrace
\begin{array}{ll}
\displaystyle{
Y=\frac{c_3-\zeta}{c_2-\zeta} \rho_2^2 - \rho_3^2.
}
\\
\displaystyle{
Y'=-\frac{c_2-c_3}{(c_2-\zeta)^2} \rho_2^2.
}
\\
\displaystyle{
Y''=-2 \frac{c_2-c_3}{(c_2-\zeta)^3}
\left[
\rho_2^2
 + \frac{c_2-\zeta}{(c_1-\zeta)(c_3-\zeta)} \rho_1 \rho_2 \rho_3 \sin \chi
\right].
}
\end{array}
\right.
\end{eqnarray}
By eliminating $\rho_j$ and $\chi$ between $Y,Y',Y''$
and the two invariants,
one builds an ODE for $Y$ \cite[Eq.~(5.42)]{MW},
which has second order, second degree and the binomial type
\begin{eqnarray}
& &
{Y''}^2=F(Y',Y,\zeta).
\end{eqnarray}
This binomial type has been ``classified'' \cite{CosScou},
i.e.~all such ODEs with the Painlev\'e property have been enumerated
and integrated.
Therefore,
if the present ODE for $Y(x)$ has the Painlev\'e property,
there should exist a homographic transformation
mapping it to one such classified ODE.
This is indeed the case, and there exists an affine transformation
\begin{eqnarray}
& &
Y=\frac{c_2-c_3}{c_2-\zeta}
\left[\lambda(\zeta) y(x) + \mu(\zeta)\right],\
x=X(\zeta),
\end{eqnarray}
which maps the ODE for $Y(\zeta)$ to the canonical ODE SD.I.a for $y(x)$
\cite[Eq.~(5.4)]{CosScou},
\begin{eqnarray}
& &
-x^2 (x-1)^2 {y''}^2 -4 y' (xy'-y)^2 + 4 {y'}^2 (xy'-y)
\nonumber
\\
& &
+A_0 {y'}^2 + A_2 (xy'-y) + \left(A_3 + \frac{A_0^2}{4}\right) y'+A_4=0.
\label{eqSDIa}
\end{eqnarray}
Among the equations determining
the three functions $\lambda(\zeta),\mu(\zeta),X(\zeta)$,
the leading ones are
\begin{eqnarray}
& &
\left\lbrace
\begin{array}{ll}
\displaystyle{
\lambda''=0,\
\mu''=0,\
\left(\lambda^2 X'\right)'=0,\
}
\\
\displaystyle{
\frac{X'}{X(X-1)} - \frac{\lambda}{(c_1-\zeta)(c_2-\zeta)(c_3-\zeta)}=0,
}
\end{array}
\right.
\end{eqnarray}
and this results in six possible values
for the three functions $\lambda(\zeta),\mu(\zeta),X(\zeta)$,
\begin{eqnarray}
& &
\lambda=(c_j-c_k)(c_l-\zeta),\
x=-\frac{(c_j-c_l)(c_k-\zeta)}{(c_k-c_j)(c_l-\zeta)},\
\end{eqnarray}
in which $(j,k,l)$ is any permutation of $(1,2,3)$.
Let us choose for instance the value
\begin{eqnarray}
& &
\lambda=-(c_3-c_1)(c_2-\zeta),
\\
& &
x=-\frac{(c_1-c_2)(c_3-\zeta)}{(c_3-c_1)(c_2-\zeta)},
\\
& &
\mu=-\frac{K_1}{2} (c_3-c_1)(c_2-\zeta)
+\frac{\beta_2}{4}
 \left[\beta_1 (c_3-c_1)(c_2-\zeta)-\beta_2 (c_1-c_2)(c_3-\zeta)\right].
\label{eqChoiceaffinetransfo}
\end{eqnarray}
The three variables $\rho_j^2$ are then linear in $y'$ and $y$,
\begin{eqnarray}
& &
{\hskip -10.0 truemm}
\rho_j^2=
\frac{(c_1-c_2)(c_2-c_3)(c_j-\zeta)}{c_2-\zeta} y'
 -(c_3-c_1)(c_j-c_2) y
\nonumber \\
& &
{\hskip -10.0 truemm}
\phantom{123}
 - (c_j-c_k)(c_j-c_l) (1-\delta_{j,2})
   \frac{K_1}{2}
 +(c_j-c_k)(c_j-c_l) \frac{\beta_2 \beta_{4-j}}{4},
\label{eqrhoj_of_y}
\end{eqnarray}
in which $(j,k,l)$ is any permutation of $(1,2,3)$,
and the link with the four constants in SD.I.a is
\begin{eqnarray}
& &
\left\lbrace
\begin{array}{ll}
\displaystyle{
A_0=- 2 K_1 - \frac{1}{2} \left(\beta_1^2+\beta_2^2+\beta_3^2\right),
}
\\
\displaystyle{
A_2= \beta_2 \left[
 \frac{\beta_3 - \beta_1}{3} K_1 + 4 K_2
 + \frac{1}{4} \beta_2^2 (\beta_3 - \beta_1)
\right],
}
\\
\displaystyle{
A_3= \beta_3 \left[
 \frac{\beta_1 - \beta_2}{3} K_1 + 4 K_2
 + \frac{1}{4} \beta_3^2 (\beta_1 - \beta_2)
\right],
}
\\
\displaystyle{
A_4= -4 K_2^2
 - \frac{5 \beta_2^2 + 2 \beta_2 \beta_3 +2 \beta_3^2}{18} K_1^2
 - \frac{2 (\beta_3-\beta_1)}{3} K_1 K_2
}
\\
\displaystyle{
\phantom{1234}
 - \beta_2^2 \frac{5 \beta_2^2 + 8 \beta_2 \beta_3 +8 \beta_3^2}{24} K_1
 - \beta_2^2 (\beta_3-\beta_1) K_2
  -\beta_2^4 \frac{\beta_2^2 + 3 \beta_2 \beta_3 +3 \beta_3^2}{16}.
}
\end{array}
\right.
\label{eqAj}
\end{eqnarray}
In order to display the permutation symmetry, it is convenient to
introduce the additional constant
\begin{eqnarray}
A_1= \beta_1 \left\lbrack
 \frac{\beta_2 - \beta_3}{3} K_1 + 4 K_2
 + \frac{1}{4} \beta_1^2 (\beta_2 - \beta_3)\right\rbrack.
\label{eqA1}
\end{eqnarray}

The equation SD.I.a,
first derived
by Chazy \cite[Eq.~B-V p.~340]{ChazyThese}
up to some homographic transformation,
has been integrated by Bureau \textit{et al.}~\cite{BGG},
and its general solution
is an algebraic transform of the generic $\PVI$ equation for $u(x)$,
\begin{eqnarray}
\PVI\ : \
u''
&=&
\frac{1}{2} \left[\frac{1}{u} + \frac{1}{u-1} + \frac{1}{u-x} \right] {u'}^2
- \left[\frac{1}{x} + \frac{1}{x-1} + \frac{1}{u-x} \right] u'
\nonumber
\\
& &
+ \frac{u (u-1) (u-x)}{x^2 (x-1)^2}
  \left[\alpha + \beta \frac{x}{u^2} + \gamma \frac{x-1}{(u-1)^2}
        + \delta \frac{x (x-1)}{(u-x)^2} \right]\ccomma
\label{eqPVI}
\\
& &
(2 \alpha,-2 \beta,2 \gamma,1-2 \delta)
=(\theta_\infty^2,\theta_0^2,\theta_1^2,\theta_x^2).
\nonumber
\end{eqnarray}
The formulae in \cite{BGG} have been further simplified
\cite[Eq.~(5.19)]{CosScou},
and the link between $\PVI$ and SD.I.a is
\begin{eqnarray}
& &
\left\lbrace
\begin{array}{ll}
\displaystyle{
y=\frac{x^2(x-1)^2}{4 u (u-1)(u-x)}
 \left\lbrace u'-\frac{u(u-1)}{x(x-1)}\right\rbrace^2
+ \frac{\Theta_\infty^2}{8} (1-2 u)
}
\\
\displaystyle{
\phantom{123}
+ \frac{\theta_0^2}{8} \left(1-2 \frac{x}{u}\right)
+ \frac{\theta_1^2}{8} \left(2 \frac{x-1}{u-1}-1\right)
+ \frac{\theta_x^2}{8} \left(1-2 \frac{x(u-1)}{u-x}\right),
}
\\
\displaystyle{
\Theta_\infty= \theta_\infty+1,
}
\\
\displaystyle{
2 A_0= \Theta_\infty^2 + \theta_0^2 + \theta_1^2 + \theta_x^2,
}
\\
\displaystyle{
4 A_1= -(\Theta_\infty^2 - \theta_0^2)(\theta_1^2 - \theta_x^2),
}
\\
\displaystyle{
4 A_2= -(\Theta_\infty^2 - \theta_x^2)(\theta_0^2 - \theta_1^2),
}
\\
\displaystyle{
4 A_3=  (\Theta_\infty^2 - \theta_1^2)(\theta_0^2 - \theta_x^2),
}
\\
\displaystyle{
32 A_4=
  (\Theta_\infty^2 + \theta_x^2)   (\theta_0^2 - \theta_1^2)^2
+ (\Theta_\infty^2 - \theta_x^2)^2 (\theta_0^2 + \theta_1^2).
}
\end{array}
\right.
\label{eqy_of_u}
\end{eqnarray}

The elimination of the intermediate constants $(A_0,A_2,A_3,A_4)$
provides the link between,
on one side the four essential parameters of the reduction
(i.e. the two first integrals $K_1,K_2$
and the three constant phases $\beta_j$ whose sum is zero),
on the other side the four monodromy exponents
$(\theta_\infty,\theta_0,\theta_1,\theta_x)$ of $\PVI$,
\begin{eqnarray}
& &
\left\lbrace
\begin{array}{ll}
\displaystyle{
4 K_1= -
\left[
\beta_1^2+\beta_2^2+\beta_3^2
+ \Theta_\infty^2 + \theta_0^2 + \theta_1^2 + \theta_x^2
\right],
}
\\
\displaystyle{
48 K_2=
 - \frac{(\Theta_\infty^2 - \theta_0^2)(\theta_1^2 - \theta_x^2)}{\beta_1}
 - \frac{(\Theta_\infty^2 - \theta_x^2)(\theta_0^2 - \theta_1^2)}{\beta_2}
 + \frac{(\Theta_\infty^2 - \theta_1^2)(\theta_0^2 - \theta_x^2)}{\beta_3}
}
\\
\displaystyle{
\phantom{1234567}
+(\beta_1-\beta_2)(\beta_2-\beta_3)(\beta_3-\beta_1),
}
\\
\displaystyle{
\beta_1 \beta_2 \beta_3 (\beta_1^2 + \beta_2^2 + \beta_3^2)
+ 2 \beta_1 \beta_2 \beta_3
   (\Theta_\infty^2 + \theta_0^2 + \theta_1^2 + \theta_x^2)
}
\\
\displaystyle{
\phantom{123}
-2 \beta_1 (\Theta_\infty^2 \theta_0^2 + \theta_1^2 \theta_x^2)
-2 \beta_2 (\Theta_\infty^2 \theta_x^2 + \theta_0^2 \theta_1^2)
-2 \beta_3 (\Theta_\infty^2 \theta_1^2 + \theta_x^2 \theta_0^2)
=0,
}
\\
\displaystyle{
-(\Theta_\infty^2 - \theta_x^2)^2 (\theta_0^2 - \theta_1^2)^2
}
\\
\displaystyle{
\phantom{123}
-2 \beta_2^2
\left[
 4 \theta_1^4 (\Theta_\infty^2 + \theta_x^2)
+4 \theta_x^4 (\theta_0^2 + \theta_1^2)
+(\Theta_\infty^2+\theta_x^2) (\theta_0^2+\theta_1^2)
 (\Theta_\infty^2+\theta_0^2-3 \theta_1^2 - 3 \theta_x^2)
\right]
}
\\
\displaystyle{
\phantom{123}
 - \beta_2^4
\left[
 (\Theta_\infty^2+\theta_0^2+\theta_1^2+\theta_x^2)^2
 -2 (\Theta_\infty^2-\theta_x^2) (\theta_0^2-\theta_1^2)
\right]
}
\\
\displaystyle{
\phantom{123}
+4 \beta_2^3 \beta_3
(\Theta_\infty^2-\theta_x^2) (\theta_0^2-\theta_1^2)
}
\\
\displaystyle{
\phantom{123}
-2 \beta_2^4 (\beta_2^2 + 2 \beta_2 \beta_3 +2 \beta_3^2)
 (\Theta_\infty^2+\theta_0^2+\theta_1^2+\theta_x^2)
}
\\
\displaystyle{
\phantom{123}
-\beta_2^4 (\beta_1^2 + \beta_3^2)^2=0.
}
\end{array}
\right.
\label{eq3waves_to_PVI}
\end{eqnarray}

The first three equations above
are invariant under both the ternary symmetry
on $\beta_j$ and the quaternary symmetry on
$(\Theta_\infty,\theta_0,\theta_1,\theta_x)$.

\section{On the singlevaluedness of the six components}
\label{sectionSinglevaluedness}

In the previous section,
we have only proven that
the variables $\rho_j^2$ and $\varphi_j'$ are single valued
(in this section, for brevity
we use ``single valued'' instead of ``with fixed critical singularities'').
However,
since the reduction is noncharacteristic,
it remains to be proven
that all the matrix elements in the reduced Lax pair (\ref{eqLax3ODE})
are also single valued,
so as to check
the conjecture of Ablowitz, Ramani and Segur \cite{ARS1980}.

This question is quite similar to a much simpler one,
which also seems to have never been investigated,
so let us first solve this question in the simple case of
the nonlinear Schr\"odinger equation,
\begin{eqnarray}
& &
i A_t + p A_{xx} + q |A|^2 A =0,\
p q \not=0,\
A \in {\mathcal C},\
(p,q)  \in {\mathcal R},\
i^2=-1.
\label{eqNLS}
\end{eqnarray}
Its traveling wave reduction
\begin{eqnarray}
& &
A(x,t)=\sqrt{M(\xi)} e^{i(\displaystyle{-\omega t + \varphi(\xi)})},\
\xi=x-ct,
\label{eqCGL3red}
\end{eqnarray}
admits the elliptic general solution
\begin{eqnarray}
& &
\left\lbrace
\begin{array}{ll}
\displaystyle{
M=-2 \frac{p}{q} \left(\wp(\xi) - \wp(a)\right),\
}
\\
\displaystyle{
\varphi' = \frac{c}{2 p}
 + \frac{j}{2} \frac{\wp'(a)}{\wp(\xi)-\wp(a)},\
j^2=-1,
}
\\
\displaystyle{
\wp(a)=(4 \omega p - c^2)/(12 p^2),\
}
\end{array}
\right.
\end{eqnarray}
in which $\wp(\xi)$ is the (even) elliptic function of Weierstrass,
and the arbitrary constants are the two elliptic invariants
$g_2, g_3$ of the function $\wp$.
Since $\wp'(a)$ is generically nonzero,
the variable $\sqrt{M}$ is multivalued
and behaves like $(\xi \pm a)^{1/2}$ near $\xi=\pm a$.
However,
the variables $e^{\pm i \arg A}$
present the same kind of branching,
so a compensation occurs making
the two fields $A$ and $\bar{A}$ singlevalued.
Indeed, the quadrature for $\varphi$ is classical
\cite[\S 18.7.3]{AbramowitzStegun},
\begin{eqnarray}
& &
\wp'(a) \int \frac{\D \xi}{\wp(\xi)-\wp(a)}
=
2 \zeta(a) \xi + \log \sigma(\xi-a) - \log \sigma(\xi+a),
\end{eqnarray}
in which the meromorphic function $\zeta$ is the primitive of $-\wp$,
the odd entire function $\sigma(z)$ behaves like $z$ near $z=0$,
and the overall expressions of $e^{i \omega t} A$
and $e^{-i \omega t} \bar{A}$ in terms of $\xi$
are indeed globally singlevalued (but not elliptic)
\begin{eqnarray}
& &
\left\lbrace
\begin{array}{ll}
\displaystyle{
e^{i \omega t} A
=
\sqrt{-\frac{2 p}{q}}
\sqrt{\wp(\xi)-\wp(a)}
\ e^{i j \zeta(a) \xi}
\left(\frac{\sigma(\xi-a)}{\sigma(\xi+a)} \right)^{i j/2}
e^{i c \xi / (2 p)} ,\
j^2=-1,
}
\\
\displaystyle{
e^{-i \omega t} \bar{A}
=
\sqrt{-\frac{2 p}{q}}
\sqrt{\wp(\xi)-\wp(a)}
\  e^{- i j \zeta(a) \xi}
\left(\frac{\sigma(\xi-a)}{\sigma(\xi+a)} \right)^{- i j/2}
e^{- i c \xi / (2 p)}.
}
\end{array}
\right.
\label{eqNLS_Traveling_wave_singlevalued}
\end{eqnarray}

Similarly, in the case of the three-wave system,
the traveling wave reduction
\begin{eqnarray}
& &
\left\lbrace
\begin{array}{ll}
\displaystyle{
    u_j(x,t)=c_j^{-1} e^{ i(\beta_j t+\alpha \xi)} \psi_j(\xi),\
}
\\
\displaystyle{
\baru_j(x,t)=c_j^{-1} e^{-i(\beta_j t+\alpha \xi)} \barpsi_j(\xi),\
}
\\
\displaystyle{
\xi=a x + b t,\
(a,b) \not=(0,0),\
\beta_1+\beta_2+\beta_3=0,
}
\\
\displaystyle{
(a c_j + b) \frac{\D}{\D \xi}
   \psi_j = i \barpsi_k \barpsi_l - i \beta_j \psi_j,
}
\\
\displaystyle{
(a c_j + b) \frac{\D}{\D \xi}
\barpsi_j=-i \psi_k \psi_l + i \beta_j \barpsi_j,
}
\end{array}
\right.
\end{eqnarray}
leads to an identical situation \cite{CRS}
\begin{eqnarray}
& &
\left\lbrace
\begin{array}{ll}
\displaystyle{
\psi_j \barpsi_j = b_j \left(\wp(\xi)-\wp(a_j)\right),
}
\\
\displaystyle{
\frac{\D}{\D \xi} \arg \psi_j =
 \hbox{constant } + \frac{j}{2} \frac{\wp'(a_j)}{\wp(\xi)-\wp(a_j)},\
j^2=-1,
}
\end{array}
\right.
\end{eqnarray}
with an identical conclusion:
singlevaluedness of $\psi_j(\xi)$ and $\barpsi_j(\xi)$.

To come back to the reduction
(\ref{eqReductionNonCharacteristic}) to $\PVI$,
establishing the singlevaluedness
of $(c_j-\zeta)^{i \beta_j} \psi_j(\zeta)$
and $(c_j-\zeta)^{-i \beta_j} \barpsi_j(\zeta)$
only requires extra care,
the result being an expression similar to
(\ref{eqNLS_Traveling_wave_singlevalued}),
in which the entire functions $\sigma(\xi-a),\sigma(\xi+a)$ of Weierstrass
are replaced by the two functions $\tau_1,\tau_2$ introduced by Painlev\'e
\begin{eqnarray}
& & u= \pm x (x-1) e^{-x} \theta_\infty^{-1} \frac{\D}{\D x}
(\Log \tau_1 - \Log \tau_2),
\end{eqnarray}
and $u(x)$ obeys the $\PVI$ equation.
These two functions $\tau_1,\tau_2$ have no movable singularities,
but they have three fixed critical singularities,
located at $x=\infty,0,1$.

\section{Dual Lax pairs for the sixth Painlev\'e equation $\PVI$}
\label{sectionDual_Lax_pairs}

The third order monodromy matrix ${\mathcal M}$
of the reduced three-wave system,
Eq.~(\ref{eqLax3ODE}),
admits in the complex $\mu$ plane
the same singularities as another third order matrix
introduced \cite{Harnad1994}
to describe the monodromy
of a time-dependent Hamiltonian with three degrees of freedom,
and later considered independently \cite{Maz2002}
from the point of view of its Laplace transform.
The common singularities of this third order monodromy matrix are
$\mu=0$ (of the Fuchsian type) and $\mu=\infty$
(of the nonFuchsian type).

Moreover, a duality has been established
by two different methods
(factorization of a residue \cite{Harnad1994},
Laplace transform in the $\mu$ space \cite{Maz2002})
between
the third order Lax pair associated with the monodromy matrix
and
a second order matrix Lax pair admitting as only singularities
four Fuchsian points.
This latter second order Lax pair indeed admits
the generic $\PVI$ equation as its zero-curvature condition.
This should have two consequences.
(i)
There should exist an identification between the two systems
(reduced three-wave, time-dependent Hamiltonian).
(ii)
The third order matrix Lax pair Eq.~(\ref{eqLax3ODE})
should have a dual, second order matrix Lax pair
admitting $\PVI$ as its zero-curvature condition.

There exists a strong motivation to have a closer look at the resulting
second order Lax pair for $\PVI$,
this is the hope that it might have a holomorphic dependence
on the four monodromy exponents $(\theta_\infty,\theta_0,\theta_1,\theta_x)$,
while the second order matrix Lax pair of Jimbo and Miwa \cite{JimboMiwaII}
has a meromorphic dependence on $\theta_\infty$.
Indeed, $\PVI$ depends holomorphically on these exponents.

Let us first review the derivation
of the second order matrix Lax pair,
then consider again the three-wave system.

\subsection{Case of the three degree of freedom Hamiltonian}

The time-dependent Hamiltonian with three degrees of freedom
\cite[Eq.~(3.56)]{Harnad1994}
\begin{eqnarray}
& &
H(q_j,p_j,x)=
 \frac{1}{4}
 \left\lbrack
 g_{31}(x) a_{13} a_{31}+g_{23}(x) a_{23} a_{32} + g_{12}(x) a_{12} a_{21}
 \right\rbrack
\label{eqHamJH}
\end{eqnarray}
with the notation
\begin{eqnarray}
& &
a_{12}=  q_1 p_2 - q_2 p_1 + (\mu_1/q_1) q_2 + (\mu_2/q_2) q_1,\
\nonumber \\ & &
a_{21}=  q_2 p_1 - q_1 p_2 + (\mu_2/q_2) q_1 + (\mu_1/q_1) q_2,\
\nonumber \\ & &
a_{13}=  q_1 p_3 + q_3 p_1 - (\mu_1/q_1) q_3 + (\mu_3/q_3) q_1,\
\nonumber \\ & &
a_{31}=  q_3 p_1 + q_1 p_3 - (\mu_3/q_3) q_1 + (\mu_1/q_1) q_3,\
\nonumber \\ & &
a_{23}=  q_2 p_3 + q_3 p_2 - (\mu_2/q_2) q_3 + (\mu_3/q_3) q_2,\
\nonumber \\ & &
a_{32}=  q_3 p_2 + q_2 p_3 - (\mu_3/q_3) q_2 + (\mu_2/q_2) q_3,
\label{eqsystemaij}
\end{eqnarray}
generates a six-dimensional first order system
made of the six Hamilton equations in the canonical variables $(q_j,p_j)$.
For the following choice of the time-dependent coefficients
\begin{eqnarray}
& &
g_{23}=\lbrack \log (g_{2}-g_{3})\rbrack ',\
g_{31}=\lbrack \log (g_{3}-g_{1})\rbrack ',\
g_{12}=0,\
(g_{2}-g_{1})'=0,\
\end{eqnarray}
this system admits
the time-independent first integral
\begin{eqnarray}
& &
I=
a_{13} a_{31}+a_{23} a_{32} + a_{12} a_{21} + 2 (\mu_1^2+\mu_2^2+\mu_3^2),
\end{eqnarray}
and the Lax pair \cite[Eqs.~(3.62), (3.65)]{Harnad1994}
(see also \cite{Maz2002}),
\begin{eqnarray}
& &
\lbrack \partial_x-{\mathcal L}_3,\partial_\lambda-{\mathcal M_3}\rbrack=0,
\nonumber \\
& & {\mathcal L}_3=
 - \lambda \pmatrix{
  g_1' & 0 & 0 \cr 0 & g_2' & 0 \cr 0 & 0 & g_3' \cr}
 + \frac{1}{2} \pmatrix{
0             & g_{12} a_{12} & g_{31} a_{13} \cr
g_{12} a_{21} & 0             & g_{23} a_{23} \cr
g_{31} a_{31} & g_{23} a_{32} &  0 \cr
}
\nonumber \\ & & \phantom{123}
 + \frac{1}{2} \diag\left(
   g_{31} \left(\mu_1 (q_3/q_1)^2 - \mu_3\right)
  +g_{12} \left(\mu_1 (q_2/q_1)^2 - \mu_2\right),
\right.
\nonumber \\ & & \phantom{123456789012} \left.
   g_{23} \left(\mu_2 (q_3/q_2)^2 - \mu_3\right)
  +g_{12} \left(\mu_2 (q_1/q_2)^2 - \mu_1\right),
\right.
\nonumber \\ & & \phantom{123456789012} \left.
   g_{31} \left(\mu_3 (q_1/q_3)^2 - \mu_1\right)
  +g_{23} \left(\mu_3 (q_2/q_3)^2 - \mu_2\right)\right),
\nonumber \\ & &
{\mathcal M}_3=
 - \pmatrix{g_1 & 0 & 0 \cr 0 & g_2 & 0 \cr 0 & 0 & g_3 \cr}
+ \frac{{\mathcal R}_{-1}}{\lambda},\
{\mathcal R}_{-1}=  \frac{1}{2} \pmatrix{
  2 \mu_1 & a_{12}  & a_{13} \cr
  a_{21}  & 2 \mu_2 & a_{23} \cr
  a_{31}  & a_{32}  & 2 \mu_3 \cr
}.
\label{eqLax3JH}
\end{eqnarray}

Let us from now on choose
the time-dependent coefficients as
\begin{eqnarray}
& &
g_{1}=0,\ g_{2}=1,\ g_{3}=x,\
g_{23}=\frac{1}{x-1},\
g_{31}=\frac{1}{x},\
g_{12}=0.
\end{eqnarray}

The residue ${\mathcal R}_{-1}$ has rank two and factorizes as
\cite{Harnad1994}
\begin{eqnarray}
& &
{\mathcal R}_{-1} = F G,\
\\
& &
F=\frac{1}{\sqrt{2}} \pmatrix{
 q_1 &  p_1 - \mu_1/q_1 \cr
 q_2 &  p_2 - \mu_2/q_2 \cr
 q_3 & -p_3 + \mu_3/q_3 \cr},\
G=\frac{1}{\sqrt{2}} \pmatrix{
  p_1 + \mu_1/q_1 &   p_2 + \mu_2/q_2 &   p_3 + \mu_3/q_3 \cr
 -q_1 & -q_2 & -q_3 \cr},
\label{eqFHFG}
\end{eqnarray}
therefore this third order matrix Lax pair $({\mathcal L}_3,{\mathcal M}_3)$
admits a dual, second order matrix
Lax pair $({\mathcal L}_2,{\mathcal M}_2)$ defined as
\cite[Eq.~(3.55), (3.61)]{Harnad1994}
\begin{eqnarray}
& &
\lbrack \partial_x-{\mathcal L}_2,\partial_\Lambda-{\mathcal M_2}\rbrack=0,
\\
& &
{\mathcal L}_2=
-\frac{R_x}{\Lambda-x},\
{\mathcal M}_2=
 \frac{R_0}{\Lambda}
+\frac{R_1}{\Lambda-1}
+\frac{R_x}{\Lambda-x},
\label{eqLax2JH}
\\
& &
R_0=- G \diag(1,0,0) F,\
R_1=- G \diag(0,1,0) F,\
R_x=- G \diag(0,0,1) F,\
\nonumber \\ & &
R_\infty=-R_0-R_1-R_x=G F,
\label{eqFourResidues}
\end{eqnarray}
with the explicit expressions for the four residues
\begin{eqnarray}
& &
2 R_\infty= \mu_1 + \mu_2 + \mu_3
\nonumber \\ & & \phantom{12}
 + \pmatrix{
 q_1 p_1 + q_2 p_2 + q_3 p_3 &
 p_1^2 + p_2 ^2 - p_3^2 -(\mu_1/q_1)^2 -(\mu_2/q_2)^2 +(\mu_3/q_3)^2 \cr
 -q_1^2 - q_2 ^2 + q_3^2 &
 -q_1 p_1 - q_2 p_2 - q_3 p_3 \cr},
\\
& &
2 R_0= \pmatrix{
 - q_1 p_1 & (\mu_1/q_1)^2 - p_1^2 \cr q_1^2 & q_1 p_1 \cr} - \mu_1,\
\\
& &
2 R_1= \pmatrix{
 - q_2 p_2 & (\mu_2/q_2)^2 - p_2^2 \cr q_2^2 & q_2 p_2 \cr} - \mu_2,\
\\
& &
2 R_x= \pmatrix{
 - q_3 p_3 & -(\mu_3/q_3)^2 + p_3^2 \cr -q_3^2 & q_3 p_3 \cr} - \mu_3.
\end{eqnarray}

The zero-curvature conditions of
$({\mathcal L}_2,{\mathcal M}_2)$ and
$({\mathcal L}_3,{\mathcal M}_3)$
are both equivalent to the Hamilton equations derived from (\ref{eqHamJH}).
Therefore, since the singularities
of the monodromy matrix ${\mathcal M}_2$
in the complex plane of $\Lambda$
are four Fuchsian singularities (located at $\Lambda=\infty,0,1,x$),
the Hamilton equations of (\ref{eqHamJH})
can be explicitly integrated in terms of $\PVI$ \cite{Harnad1994,Maz2002}.
In particular,
the invariants of the four residues
are constants of the motion,
\begin{eqnarray}
& &
\tr R_\infty=\mu_1+\mu_2+\mu_3,\
\tr R_0=-\mu_1,\
\tr R_1=-\mu_2,\
\tr R_x=-\mu_3,
\\
& &
\det R_\infty=-\frac{I}{4}
+\frac{1}{2}(\mu_1+\mu_2+\mu_3)^2,
\det R_0=\det R_1=\det R_x=0.
\end{eqnarray}

The integration of the Hamilton equations is finally performed
by identifying the coefficients of the matrix Lax pair
(\ref{eqLax2JH})
with the respective coefficients of a matrix Lax pair
for the $\PVI$ equation (\ref{eqPVI}).
If one chooses for this Lax pair the one in Ref.~\cite{JimboMiwaII},
the result is
\cite{Harnad1994}
\begin{eqnarray}
& &
\left\lbrace
\begin{array}{ll}
\displaystyle{
q_1^2+q_2^2-q_3^2=0,
}
\\
\displaystyle{
x q_1^2-\left\lbrace (1+x) q_1^2 + x q_2^2 - q_3^2\right\rbrace u=0,
}
\\
\displaystyle{
q_1 p_1 + q_2 p_2 + q_3 p_3 - 2 a_0=0,
}
\\
\displaystyle{
p_1^2 + p_2^2 - p_3^2 -(\mu_1/q_1)^2 -(\mu_2/q_2)^2 +(\mu_3/q_3)^2=0,
}
\\
\displaystyle{
\frac{q_1 p_1}{u} + \frac{q_2 p_2}{u-1} + \frac{q_3 p_3}{u-x}
+\frac{1}{u-x} - x(x-1) u'=0,
}
\\
\displaystyle{
\frac{x(x-1)u'}{u(u-1)(u-x)}
+2 v + \frac{\mu_1}{u}+ \frac{\mu_2}{u-1}+ \frac{\mu_3}{u-x}
=0,
}
\\
\displaystyle{
\frac{x(x-1)v'}{u(u-1)(u-x)}
+\left\lbrace \frac{1}{u}+ \frac{1}{u-1}+ \frac{1}{u-x}\right\rbrace v^2
}
\\
\displaystyle{
\phantom{12345}
+\left\lbrace \frac{\mu_1  }{u-1}+\frac{\mu_1  }{u-x}
             +\frac{\mu_2  }{u-x}+\frac{\mu_2  }{u  }
             +\frac{\mu_3+1}{u  }+\frac{\mu_3+1}{u-1}
 \right\rbrace v
}
\\
\displaystyle{
\phantom{12345}
+\frac{\mu_1^2+\mu_2^2+\mu_3^2+2(\mu_1+\mu_2+\mu_3)-4 a_0^2 - 4 a_0}
      {4 u(u-1)(u-x)}
=0,
}
\\
\displaystyle{
\theta_\infty^2=(2 a_0+1)^2,\
\theta_0^2=\mu_1^2,\
\theta_1^2=\mu_2^2,\
\theta_x^2=\mu_3^2,\
I=8 a_0^2.
}
\end{array}
\right.
\label{eqJHIntegration}
\end{eqnarray}

\subsection{Case of the three-wave system}

In order to perform the identification of the
Hamiltonian system (\ref{eqHamJH})
with the three-wave reduced system
(\ref{eq3waves6dimsystem}),
several methods are possible.

A first method is to factorize
the residue ${\mathcal M}_{-1}$ into a product similar to
(\ref{eqFHFG}).
Since this residue has rank three,
one first lowers its rank to two
by applying
the transition matrix $P=\mu^a {\mathcal I}$ to the Lax pair,
in which ${\mathcal I}$ is the identity matrix
and $a$ is any of the constant roots of the characteristic polynomial
(\ref{eqcharpoly}).
The factorization of the resulting rank two matrix as
\begin{eqnarray}
& &
R={\mathcal M}_{-1} -a = F G,\
\tr {\mathcal M}_{-1}=0,
\label{eqR_3waves}
\end{eqnarray}
with $F$ a $(3,2)$ matrix and $G$ a $(2,3)$ matrix,
both of rank two,
is possible \cite[\S 3.5.4]{MehtaMatrix} but it is not unique.
In particular,
if the elements of $F$ and $G$ are restricted to rational functions of
the $R_{ij}'$s,
the resulting elements of $F$ and $G$
depend on four arbitrary functions of the $R_{ij}'$s,
with no specific direct criterium to choose them,
therefore this is probably not the good method.
However,
with the definition (\ref{eqLax2JH})--(\ref{eqFourResidues}),
the invariants of the four residues
are independent of the choice of the four gauges,
\begin{eqnarray}
& &
\tr R_\infty= 3 a,\
\tr R_0=R_{11},\
\tr R_1=R_{22},\
\tr R_x=R_{33},\
\\
& &
\det R_\infty=3 a^2 + Q_2,\
\det R_0=\det R_1=\det R_x=0,
\\
& &
a^3 + Q_2 a + Q_3=0.
\end{eqnarray}

A second method consists to identify the invariants
of the two residues of the third order matrix Lax pairs.
Since the residue ${\mathcal R}_{-1}$ is not traceless,
this identification is
\begin{eqnarray}
& &
\forall z:\
\det ({\mathcal M}_{-1} - z)
=
\det \left({\mathcal R}_{-1} - \frac{\mu_1+\mu_2+\mu_3}{3}-z\right),
\end{eqnarray}
i.e.
\begin{eqnarray}
& &
\left\lbrace
\begin{array}{ll}
\displaystyle{
K_1
+\frac{\beta_1^2+\beta_2^2+\beta_3^2}{6}
=-\frac{I}{4}+\frac{(\mu_1+\mu_2+\mu_3)^2}{6},
}
\\
\displaystyle{
K_2
 -\frac{(\beta_2-\beta_3)(\beta_3-\beta_1)(\beta_1-\beta_2)}{54}
=
-i \frac{(\mu_1+\mu_2+\mu_3) I}{24}
 + \frac{5 i}{108} (\mu_1+\mu_2+\mu_3)^3.
}
\end{array}
\right.
\label{eqIdentification}
\end{eqnarray}

One difference between the two systems is the nature of the
involved constants.
The Hamiltonian system has three fixed constants
($\mu_1,\mu_2,\mu_3$)
and one movable constant (the first integral $I$),
while the reduced three-wave system has two fixed constants
(two elements among the three $\beta_j$)
and two movable constants (the two first integrals $K_1,K_2$).

\section{Conclusion}

The problem of factorizing the residue $R$ in (\ref{eqR_3waves})
is still open and currently under investigation.
If it were solved and if
the resulting second order matrix Lax pair for $\PVI$
were holomorphic in the four monodromy exponents,
this would be an improvement over
the second order matrix Lax pair of Jimbo and Miwa \cite{JimboMiwaII},
which has a meromorphic dependence on $\Theta_\infty$.
For a comparative discussion of the Lax pairs of $\PVI$,
see e.g.~Ref.~\cite{LCM2003}.

Another direction of research could be to try to match
the fourfold symmetry of $\PVI$ with the $N$-fold symmetry of the
reduced $N$-wave system.
In the case $N=3$ considered in this paper,
the correspondence between the reduced 3-wave and $\PVI$
involves
$(\Theta_\infty^2,\theta_0^2,\theta_1^2,\theta_x^2)$,
see (\ref{eq3waves_to_PVI}),
and not
$(\theta_\infty^2,\theta_0^2,\theta_1^2,\theta_x^2)$
(as is the case in e.g.~the second order scalar Lax pair of Fuchs \cite{FuchsP6}),
i.e.~it contains the shift $\Theta_\infty=\theta_\infty+1$.
This discrepancy could disappear with the four-wave system $N=4$.

Finally,
let us mention 
(and we thank the referee for signalling this reference)
a different approach \cite{KaKi} which also exhibits,
in a more general framework,
a relationship between the 3WRI and $\PVI$
and, thanks to some freedom which allows the creation of
suitable zero elements in the third order ODE Lax pair,
is able to perform a projection on a second order Lax pair
such as (\ref{eqLax2JH}).

\section*{Acknowledgments}

The authors warmly thank John Harnad and Mo Man-yue
for stimulating discussions.
This work was partially supported by
the NSERC research grant of Canada (for AMG),
the Tournesol grant no.~T2003.09 between Belgium and France (for RC and MM),
and CEA (for AMG and MM).
RC thanks the CRM for its hospitality.



\vfill \eject


\begin{thebibliography}{99}

\bibitem{AH} M.~J.~Ablowitz and R.~Haberman,
Nonlinear evolution equations--Two and three dimensions,
Phys.~Rev.~Lett.~{\bf 35} (1975) 1185--1188.

\bibitem{ARS1980} M.~J.~Ablowitz, A.~Ramani and H.~Segur,
A connection between nonlinear evolution equations and ordinary differential
equations of P-type.
{\it J.~Math.~Phys.}~{\bf 21} (1980) 715--721, 1006--1015.

\bibitem{AbramowitzStegun} M.~Abramowitz and I.~Stegun,
\textit{Handbook of Mathematical Functions},
Tenth printing (Dover, New York, 1972).

\bibitem{BGG} F.~J.~Bureau, A.~Garcet et J.~Goffar,
Transform\'ees alg\'ebriques des \'equations du second ordre dont
l'int\'egrale g\'en\'erale est \`a points critiques fixes,
Annali di Matematica pura ed applicata {\bf XCII} (1972) 177--191.

\bibitem{ChazyThese} J.~Chazy,
Sur les \'equations diff\'erentielles du troisi\`eme ordre et d'ordre
sup\'erieur dont l'int\'egrale g\'en\'erale a ses points critiques fixes,
Th\`ese, Paris (1910); 
Acta Math.~{\bf 34} (1911) 317--385.

\bibitem{Cargese1996Conte} R.~Conte,
The Painlev\'e approach to nonlinear ordinary differential equations,
{\it The Painlev\'e property, one century later},
77--180,
ed.~R.~Conte,
CRM series in mathematical physics (Springer, New York, 1999).
http://arXiv.org/abs/solv-int/9710020

\bibitem{CRS} B.~Coppi, M.N.~Rosenbluth and R.N.~Sudan,
Nonlinear interactions of positive and negative energy modes
in rarefied plasmas,
Annals of Physics {\bf 55} (1969) 207--247.

\bibitem{CosScou} C.~M.~Cosgrove and G.~Scoufis,
Painlev\'e classification of a class of differential equations of the second
order and second degree,
\SAM{\bf 88} (1993) 25--87.

\bibitem{FLMS} A.~Fokas, R.~A.~Leo, L.~Martina, and G.~Soliani,       
The scaling reduction of the three-wave resonant system and the Painlev\'e VI
equation,
Phys.~Lett.~A {\bf 115} (1986) 329--332.

\bibitem{FuchsP6} R.~Fuchs,
Sur quelques \'equations diff\'erentielles lin\'eaires du second ordre,
\CRAS\ {\bf 141} (1905) 555--558.

\bibitem{GR1998} C.~R.~Gilson and M.~C.~Ratter,
Three-dimensional three-wave interactions: A bilinear approach,
J.~Phys.~A {\bf 31} (1998) 349--367.

\bibitem{Harnad1994} J.~Harnad,
Dual isomonodromic deformations and moment maps to loop algebras,
Commun.~Math.~Phys.~{\bf 166} (1994) 337--365.

\bibitem{JimboMiwaII} M.~Jimbo and T.~Miwa,
Monodromy preserving deformations of linear ordinary differential equations
with rational coefficients.~II,
Physica D {\bf 2} (1981) 407--448.

\bibitem{KaKi} S.~Kakei and T.~Kikuchi,
The sixth Painlev\'e equation as similarity reduction of GL3 hierarchy,
16 pages, preprint http://arXiv.org/abs/nlin.SI/0508021 (2005).

\bibitem{Kitaev} A.~V.~Kitaev,
On similarity reductions of the three-wave resonant system to the
Painlev\'e equations,
J.~Phys.~A {\bf 23} (1990) 3453--3553.

\bibitem{LCM2003} Lin Run-liang, R.~Conte and M.~Musette,
On the Lax pairs of the continuous and discrete sixth Painlev\'e equations,
J.~Nonlinear Mathematical Physics {\bf 10}, Supp.~2, 107--118 (2003).
\verb+http://www.sm.luth.se/~norbert/home_journal/10s2_9.pdf and .ps +

\bibitem{MW} L.~Martina and P.~Winternitz,
Analysis and applications of the symmetry group of the multidimensional
three-wave resonant interaction problem,
Annals of Physics {\bf 196} (1989) 231--277.

\bibitem{Maz2002} M.~Mazzocco,
Painlev\'e sixth equation as isomonodromic deformations equation
of an irregular system,
\textit{The Kowalevski property}
219--238,
CRM Proc.~Lecture Notes {\bf 32}
(Amer.~Math.~Soc., Providence, RI, 2002).

\bibitem{MehtaMatrix} M.~L.~Mehta,
{\it Matrix theory}
(Les \'editions de physique, Les Ulis, 1989).

\bibitem{WTC} J.~Weiss, M.~Tabor and G.~Carnevale,
The Painlev\'e property for partial differential equations,
J.~Math.~Phys.~{\bf 24} (1983) 522--526.

\bibitem{ZM_Nwave1973} V.E.~Zakharov and S.~V.~Manakov,
Resonant interaction of wave packets in nonlinear media,
           Pis'ma Zh.~Eksp.~Teor.~Fiz.~{\bf 18} (1973) 413--417
[English~: Soviet Physics JETP Letters {\bf 18} (1973) 243--245].

\end{thebibliography}
\end{document}